\documentclass[runningheads]{llncs}

\usepackage{eccv}

\usepackage{eccvabbrv}

\usepackage{graphicx}
\usepackage{booktabs}

\usepackage[accsupp]{axessibility}  % Improves PDF readability for those with disabilities.

\usepackage{hyperref}

\usepackage{orcidlink}

\begin{document}

\title{A Compact Implicit Neural Representation for Efficient Storage of Massive 4D Functional Magnetic Resonance Imaging} 

\titlerunning{Abbreviated paper title}

% Runzhao Yang ( Tsinghua University ) < yangrz20@mails.tsinghua.edu.cn>
% Wenxin Xiang ( Tsinghua University ) < xwx22@mails.tsinghua.edu.cn> 
% Yuxiao Cheng ( Tsinghua University ) < cyx22@mails.tsinghua.edu.cn> 
% Tingxiong Xiao ( Tsinghua University ) < xtx22@mails.tsinghua.edu.cn> 
% Jinli Suo ( Tsinghua University ) < jlsuo@tsinghua.edu.cn> 
% Qionghai Dai ( Tsinghua University ) < qhdai@tsinghua.edu.cn> 

\author{Ruoran Li\inst{1}\orcidlink{} \and
Runzhao Yang\inst{1}\orcidlink{} \and
Wenxin Xiang\inst{1}\orcidlink{} \and
Yuxiao Cheng\inst{1}\orcidlink{} \and
Tingxiong Xiao\inst{1}\orcidlink{} \and
Jinli Suo\inst{1}\orcidlink{}}

\authorrunning{Li.~Author et al.}

\institute{Tsinghua University}

\maketitle

\begin{abstract}
Functional Magnetic Resonance Imaging (fMRI) data is a widely used kind of four-dimensional biomedical data, which requires effective compression. However, fMRI compressing poses unique challenges due to its intricate temporal dynamics, low signal-to-noise ratio, and complicated underlying redundancies. This paper reports a novel compression paradigm specifically tailored for fMRI data based on Implicit Neural Representation (INR). 
The proposed approach focuses on removing the various redundancies among the time series by employing several methods, including (i) conducting spatial correlation modeling for intra-region dynamics, (ii) decomposing reusable neuronal activation patterns, and (iii) using proper initialization together with nonlinear fusion to describe the inter-region similarity. This scheme appropriately incorporates the unique features of fMRI data, and experimental results on publicly available datasets demonstrate the effectiveness of the proposed method, surpassing state-of-the-art algorithms in both conventional image quality evaluation metrics and fMRI downstream tasks. This work in this paper paves the way for sharing massive fMRI data at low bandwidth and high fidelity.
\end{abstract}

%\vspace{-2mm}
\section{Introduction}
Functional Magnetic Resonance Imaging (fMRI), as a widely available non-invasive imaging tool, has been widely used in cognitive neuroscience, clinical psychology, and psychiatry. 
As large-scale fMRI datasets continue to proliferate, coupled with the ever-growing demand in both clinical \cite{vanessenWUMinnHumanConnectome2013, dimartinoAutismBrainImaging2014, poldrackOpenfMRIOpenSharing2017} and academic domains \cite{fanChallengesBigData2014, kruseChallengesOpportunitiesBig2016, razzak2018deep}, the fMRI data constitutes a significant proportion of massive biomedical data \cite{taubeNavigationVirtualReality2013, greeneFMRIInvestigationEmotional2001,DIMOKA2011S263,logothetisWhatWeCan2008}. For efficient storage and transmission, there is an urgent necessity for a high-quality data compression paradigm tailored specifically to fMRI. 

fMRI data consists of three spatial dimensions in addition to a temporal dimension, with the temporal dimension presenting intricate neural dynamics \cite{krohn2023spatiotemporal} but having a relatively low signal-to-noise ratio \cite{DIEDRICHSEN2005624}, which poses significant challenges for conventional biomedical data compression algorithms. Moreover, distinct from natural videos or ultrasound data, the redundancies in fMRI data predominantly reside within and between temporal signals of various brain regions \cite{mckeown1998analysis, van2010exploring}, rather than adjacent frames. This unique characteristic also challenges conventional compression algorithms and calls for new compressors that make extensive use of the intrinsic structure of fMRI data. %introduces complications for conventional video data compression algorithms.

\begin{figure}[t]
\centering
\includegraphics[width=\columnwidth]{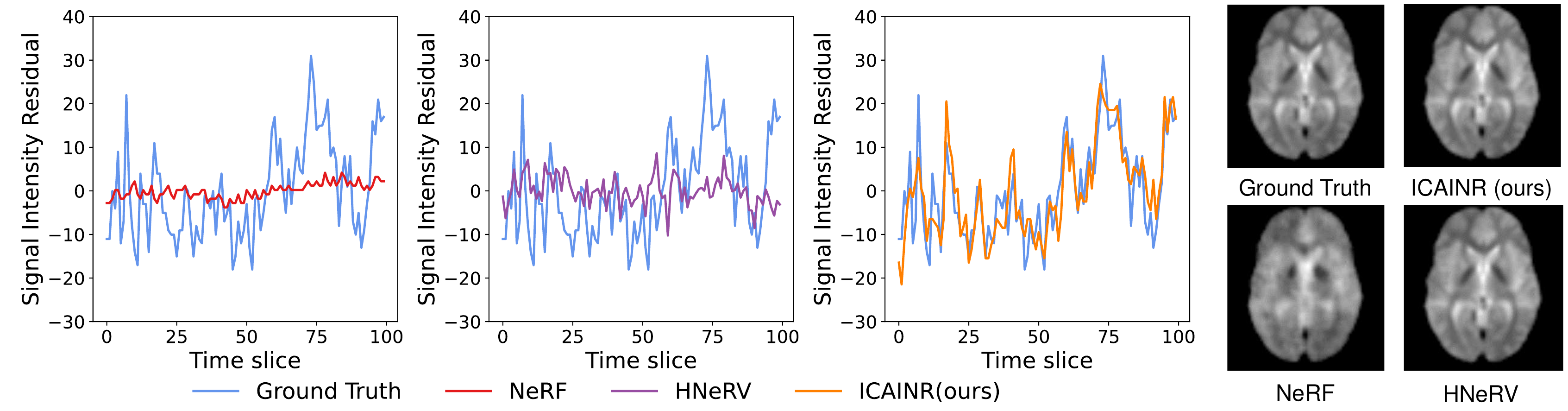}
\caption{The trace and slice comparison of an exemplar fMRI data modeled by previous INR-based methods (NeRF and HNeRV) and the proposed approach.}
\label{trace}
\end{figure}

Implicit Neural Representation (INR) is becoming increasingly widespread in various domains, such as shape representation \cite{genovaLearningShapeTemplates2019,genovaDeepStructuredImplicit2019,martel2021acorn}, scene rendering \cite{mildenhall2021nerf,zhou2021cips,wu2023hyperinr,you2023implicit}, and image/video representation \cite{skorokhodov2021adversarial,chen2022transformers,chen2021nerv,chen2023hnerv}. Due to its inherent advantages in modeling internal data correlations \cite{sitzmann2020implicit}, the recent advancements in INR-based data compression algorithms \cite{dupont2021coin,chen2021nerv,strumplerImplicitNeuralRepresentations2022,dupontCOINNeuralCompression2022,yangSciSpectrumConcentrated2023,yang2023tinc,YangBRIEF} can rival or even surpass traditional compressors, and INR stands out as the most cutting-edge and promising approach in the fields of data compression and deep learning.  
The success of INR-based compression on videos and three-dimensional biomedical data implies the feasibility but still cannot directly support its application to fMRI data. 
%As there is currently no application of INR in fMRI data compression, and considering the temporal dimension inherent in biomedical data, we will first review the applications of INR in biomedical data compression and video data compression, respectively.
First, it is straightforward but difficult to directly extend existing biomedical data compression techniques  \cite{you2023implicit} to 4D fMRI data. When directly employing INR to model the mapping from spatio-temporal coordinates to the signal values, the intricate dynamics \cite{krohn2023spatiotemporal} and heavy noise \cite{DIEDRICHSEN2005624} might result in poor encoding accuracy in the temporal dimension, as shown in Fig.~\ref{trace}. 
Second, the INR-based compression algorithms designed for conventional videos, such as HNeRV, \cite{chen2023hnerv}, focus on eliminating redundancies in adjacent video frames, and is not applicable to fMRI structure. The comparison is shown in Fig.~\ref{trace}. 
In summary, new compression techniques need to be designed to incorporate the unique features of fMRI data, including the structure, redundancy, and imaging quality.

To achieve high-quality compression of 4D fMRI data, we propose an INR-based paradigm, leveraging INR's powerful representation capability and the unique features of the fMRI data.
First, we employ a neural network to model the coordinates-to-intensity mapping, capturing the correlations of neural dynamics within the same brain region, thereby eliminating redundancies.
Next, we decompose the data into a set of reusable neuronal activation patterns and their corresponding spatial distribution, effectively removing the among-region redundancies.
Subsequently, we use INR to reduce redundancies in the spatial distribution of activation patterns and introduce a feature fusion network to simulate the integration of neuronal activation patterns in real fMRI data, enhancing the encoding precision.
Finally, we introduce an initialization scheme for neuronal activation patterns based on Independent Component Analysis (ICA) to find the primary neuronal activation patterns among the fMRI data, which can better retain the more significant activation patterns across brain regions to align the requirement of fidelity.

We demonstrate the largely superior performance of our method on publicly available datasets in order to validate it. It's worth noting that, in terms of evaluation metrics, in addition to the common metrics like PSNR and SSIM, which measure pixel-wise differences, we also introduced additional evaluation criteria as part of our assessment which are downstream tasks widely used in fMRI data analysis. Given that fMRI data is primarily utilized for data analysis rather than diagnosis observation, using the disparity between the decompressed data and the original version in downstream tasks as an additional evaluation metric allows for a more comprehensive assessment of the compression algorithms. 

In summary, the contributions of this paper can be outlined as follows:
\begin{itemize}
    \item \textbf{Four-dimensional Data Compression Based on INR.} Development of an INR-based compression paradigm tailored for the unique challenges of fMRI data.
    \item \textbf{Spatial Correlation Modeling of Local Neuronal Dynamics.} A neural-network-based approach utilizing the spatial correlation to eliminate the redundancies within local brain regions.
    \item \textbf{Decomposition and Fusion of Neuronal Activation Patterns.} A technique to describe fMRI dynamics with a nonlinear fusion of reusable activation patterns, minimizing the inter-regional redundancies. 
    % \item \textbf{Feature Fusion Network.} Implementation of a feature fusion strategy to enhance the precision of fMRI data encoding.
    %\item \textbf{ICA-Based initialization Scheme.} Adoption of an Independent Component Analysis (ICA) based representation for a compact description of neuronal dynamics. 
    \item \textbf{Initialization with Main Neuronal Activation Patterns by ICA.}  An ICA-based method to identify the dominant activation patterns in brain regions for initialization, in order to preserve salient information better. 
\end{itemize}

\begin{figure}[t]
    \centering    
    \includegraphics[width=1\textwidth]{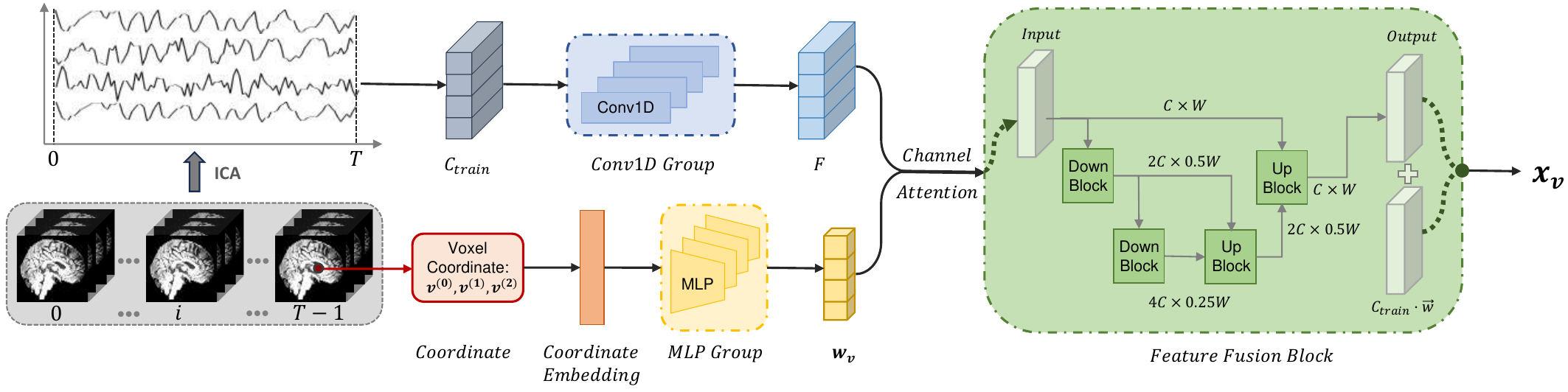}
    \caption{The workflow and the basic structure of the proposed compression approach.}
    % We firstly apply ICA to fMRI data, generating several component vectors to initialize the $C_{train}$ matrix. Then rows in $C_{train}$ will be mapped into the feature space, represented as $F$. Meanwhile, every voxel coordinate will be the input of the network, corresponding to a vector $\vec w$, which is the weight for feature channels in $F$. After Channel Attention, $F$ will be the input of Feature Fusion Block, and then generate the output $\boldsymbol{x}_{\boldsymbol{v}}$.}
    \label{method}
    % %\vspace{-4mm}
\end{figure}

\section{Related Work}
% %\vspace{-2mm}
\textbf{Data Compression.} Data compression plays a significant role in the storage and sharing of data. Over the past few decades, there has been rapid development in data compression algorithms, leading to tremendous successes with algorithms like JPEG\cite{wallaceJPEGStillPicture1992}, H.264\cite{wiegandOverview264AVC2003}, and H.265\cite{sullivanOverviewHighEfficiency2012}. The techniques which have been adopted by these compressors such as discrete wavelet transform\cite{heil1989continuous} and block-size motion compensation,have later become extensively applied in the field of data compression. In recent years, with the emergence of deep learning, data compression techniques based on deep learning have gained momentum, resulting in the development of various new compression algorithms\cite{luDvcEndtoendDeep2019, agustssonScalespaceFlowEndtoend2020}. However, when it comes to high-dimensional data, such as four-dimensional data, proper compressors are still lacking. 

%\vspace{-6mm}
\noindent\textbf{INR-based Compression.} Recent progress in data compression has introduced INR for more compact data representation, yielding promising results\cite{chen2021nerv, liENeRVExpediteNeural2022, dupont2021coin, damodaranRQATINRImprovedImplicit2023}. Unlike traditional data compression algorithms that employ explicit encoding methods, INR-based data compression algorithms leverage the powerful information capacity of neural networks to implicitly store data information within network parameters, thereby achieving efficient compression. However, for the task of fMRI compression, existing INR-based data compression algorithms have certain limitations. Some algorithms focus only on spatial or temporal redundancies, making it difficult to extend them to four-dimensional datasets or achieve satisfactory compression results. Some INR-based video compression algorithms address both types of redundancies, but due to the higher demands for fidelity in medical image compression compared to natural images, and the increased complexity of fMRI data structures, the compression strategies employed by existing INR compression algorithms are challenging to apply. In order to more effectively preserve fMRI information, we have altered the data modeling approach of INR. Instead of directly modeling fMRI signals, we utilize INR to model the interrelationships among brain region activities. %, focusing on the correlations between them.

%\vspace{-4mm}
\noindent\textbf{High Dimension Medical Data Compression.} There is a lack of adequate compression methods for four-dimensional medical data. The technical routes of these algorithms can be roughly divided into two categories. The first category is based on motion compensation, which uses motion vectors to model the difference between frames to reduce redundancies\cite{nguyenEfficientCompressionScheme2011, sanchezNovelLosslessFMRI2009, sanchezEfficient4DMotion2008, sanchezEfficientLosslessCompression2008}. However, unlike nature scene videos, the redundancies in fMRI data are concentrated between the temporal signals within and between various brain regions, rather than between adjacent frames\cite{mckeown1998analysis, van2010exploring}. Therefore, using motion compensation does not effectively utilize the spatiotemporal redundancies of fMRI for compression.
The second category is based on transform ,such as wavelet transform and so on.\cite{lalgudiFourdimensionalCompressionFMRI2005, liu2007four, rajeswari2009efficient} The transform applied to images or signals is modified to higher dimensions and then applied to medical imaging data. This modification is usually a combination of 1D transform and 3D transform, leading to limited ability to sparsely capture more complex, higher-order discontinuities\cite{BRUYLANTS2015112}.
Our method is completely different from the traditional data compression algorithms in terms of spatiotemporal redundancies compression. We leverage the correlations between brain region activities in fMRI to perform redundancies compression, fully considering the characteristics of the data.

\section{Method}
\subsection{Mathematical Representation of fMRI data}
The fMRI data $\mathbf{X}$ can be represented as a set of time series
\begin{equation}
        \mathbf{X} = \{\boldsymbol{x}_{\boldsymbol{v}} | \boldsymbol{v}^{(0)} \in [0, W), \boldsymbol{v}^{(1)} \in [0, H), \boldsymbol{v}^{(2)} \in [0, D)\}.
\end{equation}
Here $\boldsymbol{v} = [\boldsymbol{v}^{(0)}, \boldsymbol{v}^{(1)}, \boldsymbol{v}^{(2)}]$ represents the 3D spatial coordinates, and $\boldsymbol{x}_{\boldsymbol{v}} \in \mathbb{R}^{T}$ represents the fMRI signal time series at the corresponding spatial location
\begin{equation}
    \boldsymbol{x}_{\boldsymbol{v}} = [\boldsymbol{x}_{\boldsymbol{v}}^{(0)}, \boldsymbol{x}_{\boldsymbol{v}}^{(1)}, \cdots, \boldsymbol{x}_{\boldsymbol{v}}^{(T-1)}],
\end{equation}
with $W, H, D, T$ respectively denoting the width, height, depth, and length of the time series in the fMRI data.

The principle of localization in brain function organization suggests that brain functions are carried out in a set of brain regions \cite{mckeown1998analysis}. This implies that the brain space can be divided into several brain regions based on function, and there are similarities in neuronal activation within each brain region. Therefore, in fMRI data, the set of time series for each brain region can be represented as
\begin{equation}
    \mathbf{A}_{i}=\{\boldsymbol{x}_{\boldsymbol{v}} | \boldsymbol{v} \in \mathbf{V}_{i}\}, i \in [0, N),
\end{equation}
in which $\mathbf{V}_{i}$ represents the set of three-dimensional coordinates for the brain region labeled as $i$, and $N$ represents the number of brain region labels. Therefore, we have
\begin{equation}
    \mathbf{X} = \bigcup^{N-1}_{i=0} \mathbf{A}_{i},
\end{equation}
implying that $V$ can be represented as the union of $\mathbf{A}_{i}$.

\subsection{Correlation Modeling with Spatial Coordinates}

The high similarities within the neural activation in a brain region mentioned above indicate that there are high correlations between $\boldsymbol{x}_{\boldsymbol{v}}$ in the same $\mathbf{V}_{i}$, and the region $\mathbf{V}_{i}$ is usually a set of continuous space in brain. In other words, if we model the fMRI data as the mapping from spatial coordinates $\boldsymbol{v}$ to the signal time series $\boldsymbol{x}_{\boldsymbol{v}}$
\begin{equation}
    F(\boldsymbol{v}) \rightarrow \boldsymbol{x}_{\boldsymbol{v}},
\end{equation}
the local correlations of the function $F$ will be considerable, which matches with the advantages of INR-based methods that has superior capability of local structure modeling. Integrating the mathematical representation of fMRI, our parameterization function for neural network is designed at a macro level as
\begin{equation}
    f(\boldsymbol{v} | \theta) \rightarrow \widetilde{\boldsymbol{x}_{\boldsymbol{v}}}.
\end{equation}
Here $\theta$ represents the network parameters, and $\widetilde{\boldsymbol{x}_{\boldsymbol{v}}}$ represents the prediction of the ground truth $\boldsymbol{x}_{\boldsymbol{v}}$. The overall structure of the network is shown in Fig.~\ref{method}. This design aligns well with the strength of INR in modeling internal correlations, and can more effectively eliminate the redundancies of time series within local brain regions.

\subsection{fMRI time-series signal Decomposition}

The connectionism principle of brain function suggests that there exist correlations within the neuronal activation patterns in certain brain regions\cite{mckeown1998analysis, van2010exploring}. Simultaneously, the generation process of fMRI time-series signals can be theoretically viewed as the output of a Linear Time-Invariant System (LTI System)\cite{boynton1996linear, boynton2012linear}:
\begin{equation}    \boldsymbol{x}_{\boldsymbol{v}}^{(t)}=H(\sum_{i=0}^{K-1} \boldsymbol{w}_{\boldsymbol{v}}^{(i)} \cdot u_{i}(t))=\sum_{i=0}^{K-1} \boldsymbol{w}_{\boldsymbol{v}}^{(i)} \cdot H(u_{i}(t)).
\end{equation}
Here, $H$ denotes the system generating fMRI signals, $u_{i}(t)$ represents the input stimulus, $\boldsymbol{w}_{\boldsymbol{v}}^{(i)}$ refers to the stimulus distribution of signal intensity, and $H(u_{i}(t))$ is the neuronal activation pattern. Therefore, theoretically, $\boldsymbol{x}_{\boldsymbol{v}}$ can be decomposed into the weighted superposition of several time series which characterize neuronal activation patterns. If we can model the neuronal activation patterns and their distributions in fMRI which are shown in Fig. \ref{ica_example}, the network can eliminate the huge redundancies in fMRI data.

\subsection{Modeling of Neuronal Activation Pattern}

Inspired by the above prior, as shown in Fig.~\ref{method}, we have established a learnable matrix $C_\text{train} \in \mathbb{R}^{K \times T}$. Each row of this learnable matrix represents a kind of reusable neuronal activation pattern, and it is mapped to the feature space through Conv1d blocks. As such, the network is capable of learning $K$ types of reusable patterns to model the neuronal activation correlations between all 
$\boldsymbol{x}_{\boldsymbol{v}}$.

\subsection{Modeling of Activation Pattern Distribution and Feature Fusion}

While modeling several reusable neuronal activation patterns, we use INR groups to model their spatial distributions to further decrease the redundancies among each distribution
\begin{equation}
    g_{i}( \boldsymbol{v} | \sigma_{i}) \rightarrow \boldsymbol{w}_{\boldsymbol{v}}^{(i)}, i \in [0, K).
\end{equation}
Here function $g_{i}$ represents the $i$th INR network with $\sigma_{i}$ being its parameters, which models the distribution of the $i$th neuronal activation pattern. For every spatial coordinate $\boldsymbol{v}$, we concatenate the outputs of these $K$ INR networks into
\begin{equation}
    \boldsymbol{w}_{\boldsymbol{v}} = [\boldsymbol{w}_{\boldsymbol{v}}^{(0)}, \boldsymbol{w}_{\boldsymbol{v}}^{(1)}, \cdots, \boldsymbol{w}_{\boldsymbol{v}}^{(K-1)}].
\end{equation}
 The components of $\boldsymbol{w}_{\boldsymbol{v}} \in \mathbb{R}^{K}$ represent the spatial distribution values in coordinate $\boldsymbol{v}$ of each activation pattern. In this way, we can further eliminate the redundancies within the distributions of neuronal activities.

\begin{figure}[t]
  \centering
  \begin{minipage}{0.48\linewidth}
    \centering
    \includegraphics[width=1\linewidth]{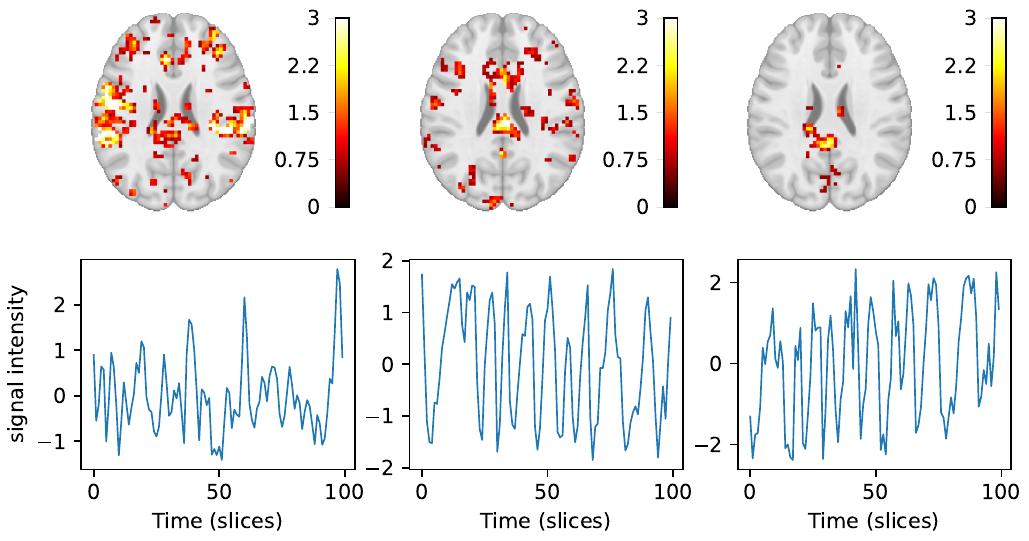}
    \caption{An example of neuronal activation patterns and their distributions.}
    \label{ica_example}
  \end{minipage}
  \hfill
  \begin{minipage}{0.48
  \linewidth}
    \centering
    \includegraphics[width=1\linewidth]{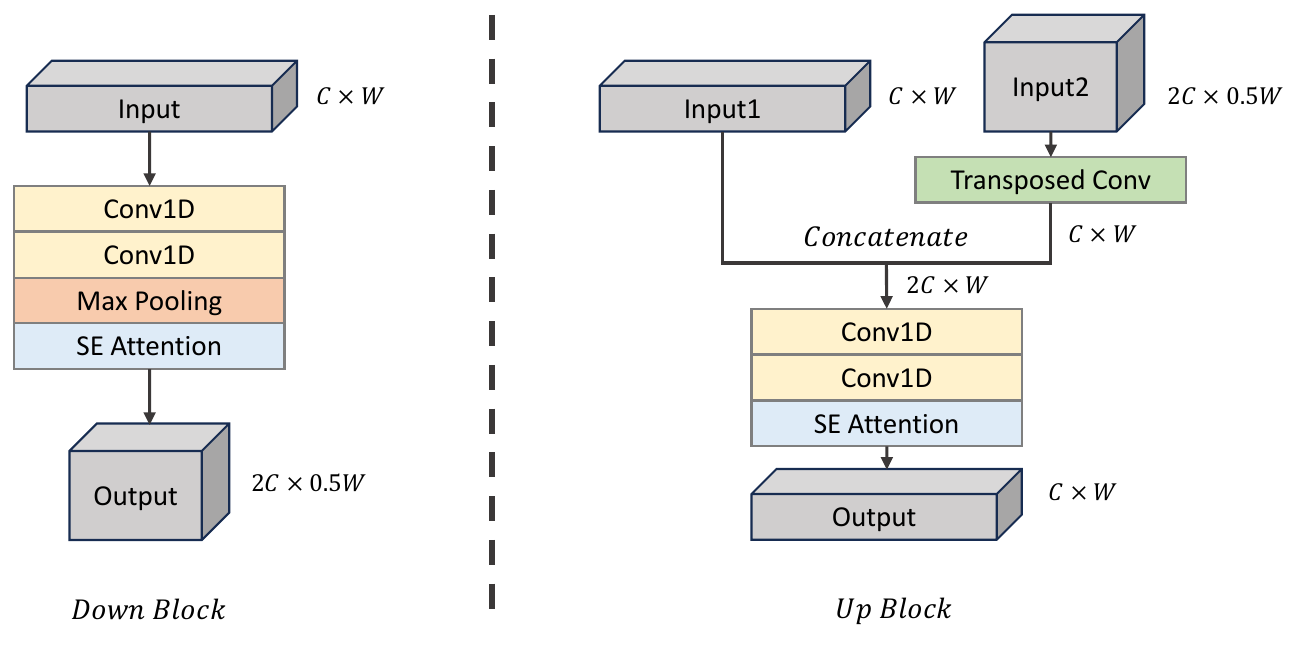}
    \caption{The network structure of the Down Block and Up Block.}
    \label{up_down_block}
  \end{minipage}
\end{figure}

Subsequently, we use the Feature Fusion Block to fuse the features of various neuronal activation patterns. This module is composed of several downsampling and upsampling operations. The internal structures of the Down Block and Up Block are shown in Fig.~\ref{up_down_block}. We concatenate the outputs of the Down Block and Up Block, which allows better utilization of the information contained in the outputs of each sampling module\cite{ronnebergerUnetConvolutionalNetworks2015}, ultimately modeling the target signal $\boldsymbol{x}_{\boldsymbol{v}}$.

\subsection{Activation Pattern Initialization by ICA}

In order to improve the convergence process of the neural network, it is crucial to design a suitable initialization for the modeling of neuronal activation patterns, i.d. matrix $C_{train}$. As described by the principle of localization\cite{mckeown1998analysis} mentioned earlier, the brain can be divided into regions, where are significant correlations among the neuronal activation patterns. 

Taking inspiration from this, we use one kind of neuronal activation pattern to describe the main neuronal activation in each certain region. Specifically, we adopted the ICA algorithm to decompose the original fMRI data into basis time series as the neuronal activation patterns, by which we initialize the learnable matrix $C_{train}$. The matrix form of ICA can be represented as $Y=AS$, where $Y$ denotes the original signal matrix decomposed into a mixing matrix $A$ and a source matrix $S=\{\vec{s}_1, \vec{s}_2, \cdots\}$. 
%These source vectors exhibit maximal independence. 
Correspondingly, applying ICA to fMRI signals gets
\begin{equation}
        \boldsymbol{x}_{\boldsymbol{v}}  = \boldsymbol{s}_{\boldsymbol{v}} \cdot \left[\boldsymbol{a}_{0}, \boldsymbol{a}_{1}, \cdots, \boldsymbol{a}_{K-1}\right]^{T}.
\end{equation}
In this equation, $\boldsymbol{x}_{\boldsymbol{v}}$ represents the fMRI time series, $\boldsymbol{a}_{i} \in \mathbb{R}^{T\times1}$ denotes the activation patterns, with the vector $\boldsymbol{s}_{\boldsymbol{v}} = [\boldsymbol{s}_{\boldsymbol{v}}^{(0)}, \boldsymbol{s}_{\boldsymbol{v}}^{(1)}, \cdots, \boldsymbol{s}_{\boldsymbol{v}}^{(K-1)}]$ describing their spatial distributions. ICA makes the independence between each distributions maximized, which allows the activation patterns we get are the main pattern in each certain region. This method make sure we can better preserve the dominant information, fulfilling the requirement of fidelity for medical data.

%The independence between each distributions is maximized because the overlap of brain region should be minimized. 

%and the spatial distribution of these main patterns should align with the corresponding brain regions they represent.

% \begin{figure}[h]
%     \centering    
%     \includegraphics[width=0.5\textwidth]{figures/ica_example.pdf}
%     \caption{The extracted regions and the ICA components of fMRI.}
%     \label{ica_example}
% \end{figure}

% \begin{figure}[t]
%     \centering    
%     \includegraphics[width=1\linewidth]{figures/up_down_block.pdf}
%     \caption{The network structure of the Down Block and Up Block.}
%     \label{up_down_block}
%     %\vspace{-4mm}
% \end{figure}

\subsection{Loss Function}
We employ a weighted combination of L2 loss and SSIM loss as the network's loss function, which is formulated as
\begin{equation}
    l = \left(1-SSIM\left(
    \{{\boldsymbol{x}_{\boldsymbol{v}}}\}
    _{\boldsymbol{v}=\boldsymbol{a}}^{\boldsymbol{v}=\boldsymbol{b}}, 
    \{\widetilde{\boldsymbol{x}_{\boldsymbol{v}}}\}
    _{\boldsymbol{v}=\boldsymbol{a}}^{\boldsymbol{v}=\boldsymbol{b}}
    \right)\right)*\sigma
    +L_2\left(
    \{{\boldsymbol{x}_{\boldsymbol{v}}}\}
    _{\boldsymbol{v}=\boldsymbol{a}}^{\boldsymbol{v}=\boldsymbol{b}}, 
    \{\widetilde{\boldsymbol{x}_{\boldsymbol{v}}}\}
    _{\boldsymbol{v}=\boldsymbol{a}}^{\boldsymbol{v}=\boldsymbol{b}}
    \right)
\end{equation}

The $\{{\boldsymbol{x}_{\boldsymbol{v}}}\}_{\boldsymbol{v}=\boldsymbol{a}}^{\boldsymbol{v}=\boldsymbol{b}}$ is a batch of ground truth data, and $\{\widetilde{\boldsymbol{x}_{\boldsymbol{v}}}\}_{\boldsymbol{v}=\boldsymbol{a}}^{\boldsymbol{v}=\boldsymbol{b}}$ is the prediction of this batch of data given by the network. $\sigma$ is a parameter to adjust L2 loss and SSIM loss at the same magnitude.

\subsection{Overview of Network's Workflow}

Based on the above analysis, we provide a summary of the workflow of the network, as shown in Fig.~\ref{method}. We apply ICA and train an individual network for each fMRI data to get higher fidelity. Firstly, we calculate the mean frame of the fMRI data and subtract it from each frame. We will store the mean frame individually utilizing conventional compression like PNG due to the high structure similarities between frames. Before training, we apply ICA to the fMRI data in which the black areas are masked. The ICA results will be used to initialize $C_{train}$ and pre-train the network. When the training process begins, every row vector in $C_{train}$ will be the input of $K$ Conv1d blocks respectively, and the output will be concatenated into a feature map $F$. Meanwhile, the coordinate $\boldsymbol{v}$ will be embedded to serve as the input of $K$ MLPs, and the embedding method we leverage is the same one in NeRF\cite{mildenhallNerfRepresentingScenes2021a}. The output of MLPs are single weight numbers $\boldsymbol{w}^{(i)}$ which are concatenated into the weight vector $\boldsymbol{w}$. During the process of Channel Attention, $\boldsymbol{w}$ and $F$ will be integrated and serve as the input of the Feature Fusion Block. The Feature Fusion Block contains several Down Blocks and Up Blocks respectively for downsampling and upsampling, whose structures are shown in Fig.~\ref{up_down_block}. Finally, a Conv1D block will be applied to the output of Feature Fusion Block and generate the output vector $\boldsymbol{x}_{\boldsymbol{v}}$. 

\subsection{Model Compression}
Model pruning, quantization, and entropy coding are prominent techniques in model compression\cite{hanDeepCompressionCompressing2016}.To speed up the training process, we do not apply model pruning in our network. Upon completion of training, we proceed to quantize the network parameters and subsequently utilize Huffman coding as the entropy coding method.

% \begin{figure}[t]
%     \centering    
%     \includegraphics[width=\linewidth]{figures/trace.pdf}
%     \caption{The four traces of all the methods. }
%     \label{tracek}
%     %\vspace{-4mm}
% \end{figure}

\section{Experiments and Analysis}
\subsection{Implementation Details}
We conducted experiments on four fMRI datasets collected for different downstream tasks:
Three datasets are from OpenfMRI\footnote{These three datasets were obtained from the OpenfMRI database, with accession numbers  ds000007, ds000101, ds00102, respectively.}, an open-source repository for the free and open sharing of fMRI datasets. The fourth dataset is the widely used Haxby dataset, a pioneering study of brain pattern recognition\cite{haxbyDistributedOverlappingRepresentations2001}, which has extended time series and suitable for the fMRI classification task.

During the training stage, we randomly selected fMRI data from the four datasets. For the first three datasets, in order to facilitate subsequent analysis, we uniformly preprocessed the fMRI data by aligning it to a standard brain template. This resulted in voxel dimensions of 64$\times$64$\times$48 and a time series length of 100. The Haxby dataset comes with a pre-matched mask. And we did not perform registration, as it is unnecessary for the classification task. Due to its longer time series, we addressed CUDA memory limitations by slicing the data in Haxby.

In the experiments, our MLP network was configured with five layers, and the frequency of the coordinate embedding was set to 10. We used JPEG to compress the mean of the data, and utilized the Adamax optimizer with an initial learning rate of 8e-4. The training epoch was set to 1500. We also utilized ICA's output network pre-training. The compression ratio of the network can be adjusted by varying parameters such as the number of ICA components, average frame compression quality, count of the MLP parameters, number of convolution channels and network layers. All these parameters can be adjusted using stored YAML files.

% \begin{figure}[t]
%     \centering
%     \begin{minipage}{0.766\textwidth}
%         \includegraphics[width=\linewidth]{figures/vision.pdf}
%      \end{minipage}
%     \caption{The fMRI slices from the ground truth and decompressed versions of different algorithms. Here the size of the whole fMRI data volume is 64 $\times$ 64 lateral pixels, 48 layers, and 100 frames, while the slices we picked are the central slice. }
%     \label{slice}
% \end{figure}

\begin{figure}[t]
    \centering
    \includegraphics[width=1\linewidth]{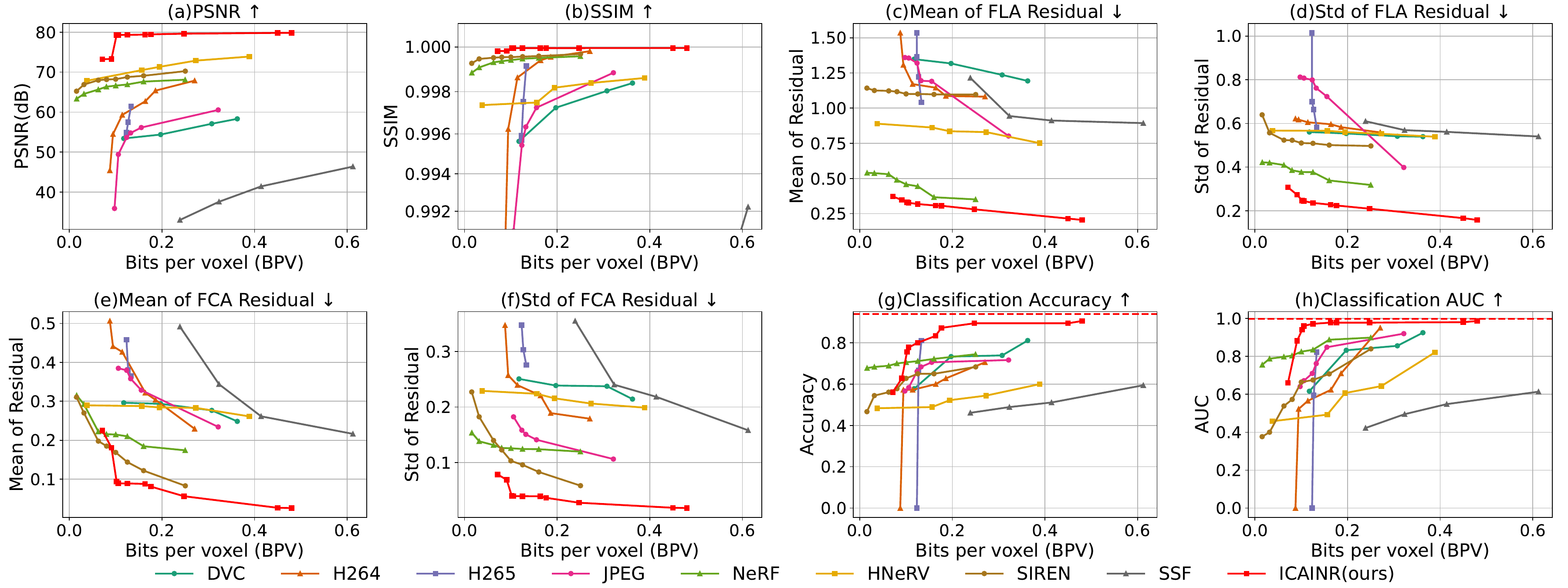}
    \caption{The performance of the proposed approach and the benchmark algorithms.}
    \label{all_metrics}
    %\vspace{-4mm}
\end{figure}

\begin{figure}[t]
    \centering    
    \includegraphics[width=\linewidth]{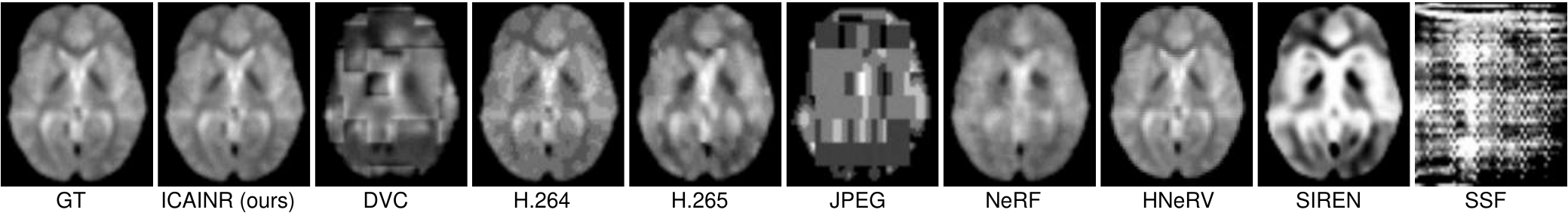}
    %\vspace{-4mm}
    \caption{The fMRI slices from the ground truth and decompressed versions of different algorithms. Here the size of the whole fMRI data volume is 64 $\times$ 64 lateral pixels, 48 layers, and 100 frames, while the slices we picked are the central slice.}
    \label{slice}
\end{figure}

\begin{table*}[t]
    \scriptsize
    \centering
    \caption{The performance of different compressors at a specific compression ratio, roughly 100$\times$.Methods marked in {\color{red}red} and with the * suffix are the best performing methods, while those marked in {\color{blue}blue} and with the \# suffix are the second best performing methods.}
    %\vspace{-2mm}
    \resizebox{\textwidth}{!}{
    % \fontsize{8pt}{9.7pt}\selectfont
    \begin{tabular}[t]{c|c|c|c|c|c|c|c|c|c} 
    \hline
    Method &  ICAINR(ours)  & H.264 & H.265 & JPEG & NeRF & HNeRV & SIREN & SSF& DVC\\\hline
    Compression Ratio$\uparrow$ & {\color{red}~127.83$\times$*}&97.49$\times$ & {\color{blue}125.99$\times$\#} & 102.95$\times$ &99.77$\times$ &102.71$\times$ &99.76$\times$ &~66.99$\times$ &81.13$\times$ \\\hline
    PSNR(dB)$\uparrow$ & {\color{red}79.31*} &65.37 & 61.49 & 56.14 & 67.63 & {\color{blue}70.49\#} & 69.09 & 33.03  & 57.09\\\hline
    1 - SSIM$\downarrow$ & {\color{red}5.54E-5*}  &	1.96E-3 & 8.38E-4 & 2.74E-3 & 4.83E-4  & 2.62E-3  & {\color{blue}4.01E-4\#} & 9.39E-2    & 4.39E-4\\\hline
    Mean of FLA Residual$\downarrow$ & {\color{red}0.32*} & 1.15 & 1.04 & 1.19 & {\color{blue}0.37\#} & 0.86 & 1.10 & 1.21  & 1.32 \\\hline
    Std of FLA Residual$\downarrow$ & {\color{red}0.24*} & 0.60 & 0.58 & 0.72 & {\color{blue}0.34\#} & 0.57 & 0.50 & 0.61  & 0.55\\\hline
    Mean of FCA Residual$\downarrow$ & {\color{red}0.09*} & 0.32 & 0.37 & 0.33 & 0.18 & 0.29 & {\color{blue}0.12\#} & 0.50 & 0.29 \\\hline
    Std of FCA Residual$\downarrow$ & {\color{red}0.04*} & 0.22 & 0.28 & 0.14 & 0.12 & 0.22 & {\color{blue}0.08\#} & 0.35 & 0.24\\\hline
    \end{tabular}
    }
    \label{specified_ratio}
\end{table*}

\subsection{Benchmark Methods and Evaluation Metrics}
To demonstrate the advantages of the proposed approach, we comprehensively compare it with eight state-of-the-art (SOTA) algorithms, which can be broadly categorized into three groups: widely used traditional compression algorithms like JPEG\cite{wallaceJPEGStillPicture1992}, H.264\cite{wiegandOverview264AVC2003}, and H.265 (HEVC)\cite{sullivanOverviewHighEfficiency2012}, data-driven deep learning compression algorithms such as SSF\cite{agustssonScalespaceFlowEndtoend2020} and DVC\cite{luDvcEndtoendDeep2019}, and implicit neural representation-based compressors, including HNeRV\cite{chen2023hnerv}, NeRF\cite{mildenhallNerfRepresentingScenes2021a}, and SIREN\cite{sitzmann2020implicit}. For the competitors designed for 2D or 3D data, we compressed the 4D fMRI data slice by slice or concatenated the slices into 2D or 3D data.

We utilized the OpenCV implementation of JPEG and the FFmpeg implementation of H.264 and H.265. We set the compression ratio by calculating the corresponding bit rate. For DVC and SSF, which are data-driven methods, we changed the compression ratio by specifying the quality parameters and fine-tuning the pre-trained models provided by their authors. For SIREN and NeRF, we set the layer numbers of MLP to 7, and for HNeRV, we specified the network parameters to achieve different compression ratios.

The evaluation metrics employed in our experiments can be divided into two parts. The first part involves traditional image quality evaluation metrics PSNR and SSIM, which are compared across various compression ratios. The second part pertains to downstream tasks based on fMRI data. We selected three downstream tasks, which are all classical and widely used methods in fMRI analysis to comprehensively evaluate the compression quality and fidelity. These tasks
quantitatively measure the preservation of various informative cues within the data, including the information about stimuli-brain, intra-brain, and fMRI signal patterns, covering a significant portion of fMRI analysis.
These downstream tasks encompass the following: 

\noindent i) General linear model First Level Analysis (FLA): This task involves fitting and hypothesis testing of fMRI waveforms to compute the strength of association between specific stimuli/tasks and various brain regions\cite{friston1994statistical}, which is the most widely used in fMRI statistical parametric mapping\cite{smith2004overview}. We conduct FLA on three datasets and use the mean absolute difference between the statistical maps obtained from the original and compressed data for model evaluation.

\noindent ii) Brain regions Functional Connectivity Analysis (FCA): The brain connectome characteristics have offered valuable insights to explain the diversity of pathological conditions and behaviors across different peoples\cite{mohanty2020rethinking}, and functional connectivity is one of the most widely used connectome characteristics. It is defined as the correlation coefficients between the voxel waveforms of different brain areas\cite{salvador2005neurophysiological}. In this task, the same dataset was used as in the FLA task, with the MSDL atlas template provided in nilearn\cite{varoquauxMultisubjectDictionaryLearning2011}. Again, we use the mean absolute difference between the correlation matrices from the ground truth and decompressed fMRI for performance evaluation. 

\noindent iii) fMRI-based Classification Task (CT): Decoding or pattern recognition techniques are a significant part of fMRI analysis\cite{haxbyDistributedOverlappingRepresentations2001}. With the development of machine learning, there has been growing interest in decoding fMRI data by machine learning. We use a linear classifier based on SVM\cite{de2008combining}, which is most commonly used in fMRI classification\cite{naselaris2011encoding}, to do our experiment. In this task, we employed the Haxby dataset with various kinds of images serving as stimuli. We trained the SVM  with the voxel waves, to distinguish %. Specifically, we conducted a binary classification between 
stimuli images of the house and face. %The input of the classifer is voxel values of one fMRI frame and the output is the probability of the stimuli type. 
We applied 10-fold cross-validation and used the cross-validation classification accuracy of the decompressed fMRI as the evaluation metric.

\subsection{Experiment Results}
\textbf{PSNR \& SSIM.}We first validate the data fidelity after fMRI compression, and compare our approach against existing compression algorithms. Here we use PSNR and SSIM as quantitative evaluation metrics, and provide the scores of different algorithms across varying compression ratios, as plotted in Fig.~\ref{all_metrics}{\color{red}a} and Fig.~\ref{all_metrics}{\color{red}b}. Notably, under similar compression ratios, our method consistently outperforms existing SOTA algorithms at both higher and lower compression ratios. When changing the compression ratio, some algorithms fluctuate, in contrast to the remarkably stable performance of our approach. 

\begin{figure}[t]
    \centering
    \includegraphics[width=1\textwidth]{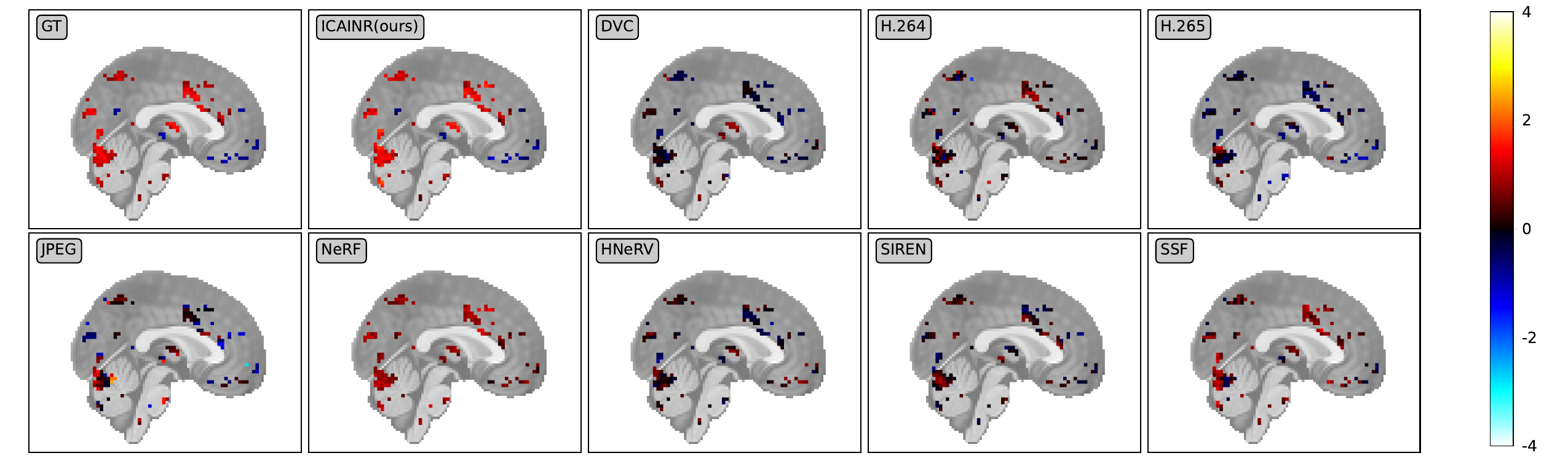}
    \caption{The brain t-score map calculated from the ground truth and decompressed results by different algorithms. We set the confidence level to be 90\% to label the voxels which are highly correlated with the stimuli. Here the displayed intersecting plane of the brain map is the center frame along $y$ axis.}
    \label{first_level_brain}
    %\vspace{-2mm}
\end{figure}

%\vspace{-5mm}
\noindent\textbf{Visual Results.}The visual comparison of the decompressed data by various algorithms at a fixed compression ratio ($\sim$100$\times$) is shown in Fig.~\ref{slice}. Note that the actual achieved compression ratio of different algorithms differs slightly, as shown in Tab.~\ref{specified_ratio}, because one cannot specify the final compression ratio exactly. From the results, it is evident that despite yielding relatively high PSNR and SSIM values, many algorithms failed to deliver satisfactory visual quality. For instance, JPEG and DVC suffer from block effect\cite{lee1998blocking}, with noticeable fragmentation between adjacent image blocks. The INR-based algorithms, like NeRF, SIREN, and HNeRV, sacrificed a considerable amount of high-frequency details and thus show over-smoothness. In contrast, other algorithms such as H.264 and H.265 exhibited noticeable noise, and it seems that SSF almost fails to model the fMRI data.

% Moreover, we can notice that multiple algorithms of low visual quality achieve extremely high PSNR, but not the case for the fMRI data. %because most of the decompressed results' PSNR higher than 50dB, which is considered as a representative of high image quality. 
% This observation underscores the importance of introducing metrics other than PSNR and SSIM to evaluate the effectiveness of data compression algorithms for fMRI. Consequently, we introduce several typical downstream tasks on fMRI data as additional evaluation metrics to provide a more comprehensive assessment of compression algorithms. Across different compression ratios, we conducted FLA, FCA, and CT on the decompressed data obtained from various algorithms. 

\begin{figure}[t]
    \centering    \includegraphics[width=1\textwidth]{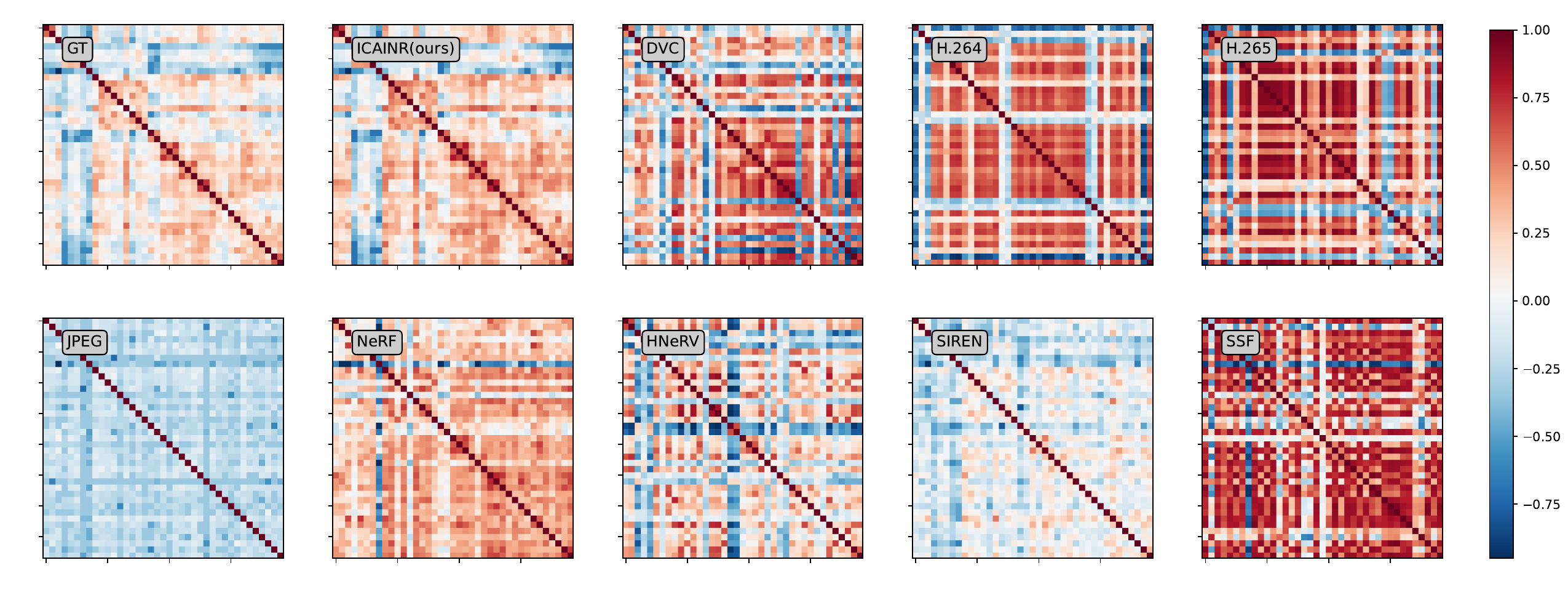}
    %\vspace{-1mm}
    \caption{The brain region connectivity matrix involving 39 brain regions calculated from ground truth and decompressed results of all algorithms. Each matrix entry is calculated as the correlation coefficients between two corresponding regions. %We do not show the brain region labels along x and y axis because of the high label density.
    }
    \label{connectivity_brain}
    %\vspace{-1mm}
\end{figure}

%\vspace{-5mm}
\noindent\textbf{FLA Results.}To evaluate the behaviors of our compressor and other competitors on FLA, we computed the mean and standard deviation of residue between the compressed fMRI data and ground truth FLA results, as plotted in Fig.~\ref{all_metrics}{\color{red}c} and Fig.~\ref{all_metrics}{\color{red}d}. The curves indicate that our method has the smallest residue and performance fluctuation under all compression ratios among these algorithms. 
Among other algorithms, NeRF can achieve results slightly inferior to ours by around 0.1 and 0.15 in terms of mean and standard deviation on average, while the performance of the remaining competitors is quite limited. 
Additionally, we display the visualized FLA obtained by various algorithms at a compression ratio about  $\sim$100$\times$, as illustrated in Fig.~\ref{first_level_brain}. We can observe that our method exhibits high similarity to the ground truth, while others show significant differences. These results indicate that our method effectively preserves the activities caused by stimuli, which are one of the most important components of fMRI signals.

%\vspace{-5mm}
\noindent\textbf{FCA Results.}Regarding to the influence of compression on FCA, similar to FLA, we illustrated the mean and standard deviation of residual after decompression in Fig.~\ref{all_metrics}{\color{red}e} and Fig.~\ref{all_metrics}{\color{red}f}. The curves show that our method has much less information loss than other algorithms if the compression ratio is lower than 170$\times$, and the average loss approximates zero when compression ratio is around 30$\sim$40. 
When compression ratio is higher, the performance in mean residual of NeRF, SIREN, and our method is similar. From the the standard deviation, one can conclude that our approach exhibits the lowest performance fluctuation among all the algorithms working at a similar compression ratio, and achieves stable performance (standard deviation close to 0) when the compression ratio is lower than 100$\times$. 
We also compare the visualized brain connectivity at a compression ratio of about 100$\times$ in Fig.~\ref{connectivity_brain}, from which one can observe that our compressor presents the highest similarity to ground truth. These results demonstrate our capability to preserve the correlation information among brain regions.

%\vspace{-4mm}
\noindent\textbf{CT Results.}To test the effectiveness of our compressor and its advantage over previous competitors in the successive classification task (CT),  we calculated the 10-fold cross validation accuracy and AUC of the decompressed data compared to the original data before compression, as shown in Fig.~\ref{all_metrics}{\color{red}g} and Fig.~\ref{all_metrics}{\color{red}h}. The results demonstrate that our algorithm outperforms other methods at compression ratio lower than 170, and can achieve accuracy and AUC close to the ground truth, 93.89\% and 0.9975 respectively, when the compression ratio falls below 100$\times$. In other words, our compression preserves most of the distinctive features in the fMRI data volumes.

% \begin{figure}[h]
%     \centering
%     \begin{minipage}{0.495\linewidth}
%         \includegraphics[width=1\linewidth]{figures/classification_accuracy.pdf}
%     \end{minipage}
%     \begin{minipage}{0.495\linewidth}
%         \includegraphics[width=1\linewidth]{figures/classification_auc.pdf}
%     \end{minipage}
%     \caption{The classification accuracy of the recorded FMRI data ad that of the compressed counterparts by different compression algorithms. % for ground truth and the decompressed results of all algorithms. 
%     The dotted line plots the classification accuracy achieved on the original recorded data, while the other solid lines correspond to the results on different compressed versions.}
%     \label{acc_auc}
% \end{figure}

In summary, our algorithm showcases superior outcomes in terms of both PSNR, SSIM, as well as impressive score in downstream tasks, 
%Particularly notable is our algorithm's impressive performance in downstream tasks, where the demand for compression fidelity is notably high.
while some previous algorithms focus on PSNR and SSIM but suffer from degraded performance in successive analysis.  
This robust performance substantiates our algorithm's capacity to effectively retain crucial information in fMRI, providing the necessity to take into account the physical meaning of fMRI data during compression and presenting the potential to perform processing and analysis on fMRI data at low bandwidth. % get a relatively high compression ratio with acceptable loss in application.

\begin{table*}[t]
\scriptsize
   \centering
   \caption{All methods used in ablation study. Methods marked in {\color{red}red} and with the * suffix are the best performing methods, while those marked in {\color{blue}blue} and with the \# suffix are the second best performing methods.}
   %\vspace{-2mm}
   \resizebox{\linewidth}{!}{
   % \fontsize{4.7pt}{5.6pt}\selectfont
   \begin{tabular}[t]{c|c|c|c|c|c} 
   \hline
    Method & Compression Ratio$\uparrow$ & PSNR(dB)$\uparrow$ & 1 - SSIM$\downarrow$ & FLA Residual$\downarrow$ & FCA Residual$\downarrow$ \\\hline
    ICAINR(origin) & {\color{red}173.14$\times$*} & {\color{red}78.74*} & {\color{red}2.28E-5*} & {\color{red}0.32$\pm$0.24*} & {\color{red}0.14$\pm$0.10*}\\\hline
    ICAINR(without fusion) & {\color{blue}172.96$\times$\#} & {\color{blue}78.67\#} & {\color{blue}2.31E-5\#} & {\color{blue}0.33$\pm$0.25\#} & {\color{red}0.14$\pm$0.10*}\\\hline
    ICAINR(uniform init) & 164.96$\times$ & 78.60 & 2.36E-5 & 0.36$\pm$0.30 & 0.28$\pm$0.09\\\hline
    ICAINR(normal init) & 165.33$\times$ & 78.17 & 2.56E-5 & 0.49$\pm$0.35 & 0.22$\pm$0.07\\\hline
    SIREN(3D) & 160.51$\times$ & 72.93 & 6.06E-5 & 0.36$\pm$0.33 & 0.50$\pm$0.16\\\hline
    SIREN(4D) & 159.06$\times$ & 70.77 & 1.04E-4 & 1.11$\pm$0.51 & {\color{blue}0.18$\pm$ 0.11\#}\\\hline
   \end{tabular}
   }
\label{ablation}
\end{table*}

\subsection{Ablation Studies}
In this section, we quantitatively determine the contribution of the key modules in the proposed approach. %  the effectiveness of our network design through a comparison between several methods. The results obtained are shown in the Tab.~\ref{ablation}.

% %\vspace{-4mm}

% \textbf{Correlation Modeling with Spatial Coordinates.}
% To prove the superiority of using INR to model fMRI data with 3D spatial coordinates, we compared results from SIREN which uses 3D spatial coordinates and 4D spatiotemporal coordinates as input, under the approximate compression rate(row 5 and row 6 in Tab.~\ref{ablation}). The results in the table show that the SIREN model using spatial coordinates performed better, and the 3D coordinates can better model the correlation of fMRI time series.

%\vspace{-4mm}
\noindent\textbf{Decomposition of Neuronal Activation Patterns.}
We compared SIREN, an INR-based method modeling the data directly, with our method. We removed the Feature Fusion Block in our network to eliminate its impact. As listed in 2nd, 5th and 6th rows of Tab.~\ref{ablation}), which showed better performance in terms of all metrics. The results show that using pattern decomposition is advantageous over directly modeling the 4D fMRI data with INR, given the high complexity of fMRI signals and various underlying correlations.

%\vspace{-4mm}
\noindent\textbf{Feature Fusion Block.}
At a similar compression ratio, we compare the quality of the decompressed data before and after removing the feature fusion block. The results are shown in the  1st and 2nd rows of Tab.~\ref{ablation}, demonstrating that the Feature Fusion Block achieves higher data fidelity of the target fMRI signals than naive linear modeling. % than modeling the data linearly.

%\vspace{-4mm}
\noindent\textbf{ICA Initialization.}
To verify the effectiveness of ICA initialization, we compared the performance with the $C_\text{train}$ matrix uses ICA initialization, Uniform Initialization, and Normal Initialization 1st, 3rd and 4th rows in Tab.~\ref{ablation}). The results, especially the performance in downstream tasks, demonstrate that ICA initialization can enhance the network's ability to model the signals, especially the neural activation information and correlations.

\section{Summary and Discussions}
In this paper, we introduce a novel INR-based compression paradigm for fMRI data. By leveraging the strong representation capability of deep neural networks and taking into account of fMRI's unique characteristics, we aim to compactly describe the neural activation patterns and their spatial distributions, which reduces both temporal and spatial redundancies in the raw recordings. %We employ modules such as Attention and CNN to fuse the features of these patterns, and subsequently predict the fMRI time series. 

%\vspace{-5mm}
%Our model represents the inter-regional correlations of fMRI successfully. 
\noindent\textbf{Advantages.} 
Based on the comprehensive experiments conducted, our algorithm outperforms existing SOTAs in terms of data fidelity such as PSNR and SSIM, as well as evaluation metrics for downstream fMRI tasks, which meets the fidelity requirements for medical image compression.
Moreover, our algorithm achieves a higher compression ratio and outperforms existing SOTA methods in terms of both image fidelity and downstream tasks. 
Notably, our study pioneers the application of INR-based compression methods to four-dimensional biomedical data. Additionally, we present a novel INR-based compression framework, offering valuable insights for future research in INR-based biomedical image compression.

%\vspace{-4mm}
\noindent\textbf{Limitations and Future Work.}
As a preliminary research on INR-based fMRI compression, our approach can be extended in several aspects before becoming a mature compression tool. Firstly, as a deep learning-based compression algorithm, our approach takes longer time than conventional compressors. Therefore, we are exploring novel architectures and techniques, such as meta-learning, for fast model learning. 
%In the future, we plan to for higher compression ratio and faster compression. 
%Consequently, the compression speed of our algorithm is comparatively slower than traditional image or video compression methods. 
Secondly, we use all the initial components obtained from ICA, but some of them contain little information about the neural activities. We are also working on strategies for extracting informative signals to avoid compressing unnecessary components and thus build a more compact compression. 
Thirdly, we adopt a widely used method for network compression, and there is room for designing more tailored model compression techniques for our network in the future.

\bibliographystyle{splncs04}
\bibliography{main}
\end{document}